\documentclass[preprint,aps]{revtex4}
\usepackage{amsmath,amssymb}
\usepackage{graphicx}
\usepackage{dcolumn}
\usepackage{bm}
\usepackage{epstopdf}
\begin{document}
\author{Ernesto Gonz\'alez-Candela$^{1}$ and V. Romero-Roch\'{\i}n$^2$} 
\affiliation{
$^{1}$Universidad Aut\'onoma Metropolitana - Iztapalapa. 09340 M\'exico, D.F. 
Mexico
\\
$^{2}$ 
 Instituto de F\'{\i}sica, Universidad
Nacional Aut\'onoma de M\'exico, Apdo. Postal 20-364, M\'exico D.
F. 01000, Mexico. }

\title{ Directed transport as a mechanism for protein folding in vivo}

\date{\today}
\begin{abstract}
We propose a model for protein folding \emph{in vivo} based on 
a Brownian-ratchet mechanism in the multidimensional energy
landscape space. The device is able to produce directed transport 
taking advantage of the assumed intrinsic 
asymmetric properties of the proteins and employing 
the consumption of energy provided by an external source.
Through such a directed 
transport phenomenon, the polypeptide finds the native 
state starting from any initial state in the energy landscape 
with great efficacy and robustness, even in the presence of different
type of obstacles. This model solves Levinthal's paradox without requiring 
biased transition probabilities but at the expense of opening the system
to an external field. 
\end{abstract}
\maketitle


\section{Introduction}

Following the classical work of Anfinsen \cite{Anfinsen,Anfinsen2} on 
protein folding, Levinthal \cite{Levinthal} argued that if 
the protein multidimensional free energy landscape had a ``golf course'' 
like shape, with the hole being the native state, it 
would take essentially forever for a protein to fold correctly, 
provided the search were performed randomly. This is known as Levinthal's 
paradox, since proteins \emph{in vivo} fold very fast and with high 
efficacy. However, Zwanzig \emph{et al} \cite{Zwanzig,Zwanzig2} 
demonstrated that, within a random search, a small bias for 
configurations closer to the native state would be enough 
to solve the paradox. Alternatively, to explain the efficacy 
of folding, the concept of a funneled energy 
landscape was developed 
\cite{Wolynes,Wolynes2,Onchuic,Dill,Miller,Dill2}, 
with the further possibility of folding pathways. In such an scenario the 
process is driven mainly by 
free energy differences, that is, by ``falling'' through the funnel into the 
state of lowest free energy. This idea has been central in 
the study of protein folding. In this article, we propose 
a further alternative way out to the so-called Levinthal paradox, which does not
necessarily needs a funnel structure nor introduces ad-hoc biased probabilities. 
We appeal to a Brownian motor or ratchet-like mechanism 
for the process of protein folding. This is a non-equilibrium 
process that requires, first, a free energy landscape with a 
ratchet-like asymmetry, and second, the consumption of energy 
from an external source \cite{GCRR}. We shall discuss below how 
these two essential ingredients may be justified.

The ratchet mechanism to produce directional current or 
motion has become a paradigm of the interplay of nonlinear 
phenomena with Brownian motion and unbiased external noise in 
mesoscopic systems \cite{Magnasco,Reimann,Hanggi1,Bartussek}.
We shall use the term ``ratchet'' for any conservative 
mechanical system that has an ingrained spatial asymmetry; 
a simple one-dimensional example is a particle in a 
saw-tooth like potential. Due to the potentiality
of these systems to describe mesoscopic systems such as 
biological ones  \cite{Reimann,Hanggi1,Bartussek,Astumian1},
one should consider the ratchet to be immersed in a thermal 
bath. This adds dissipative 
and stochastic thermal forces that are related to each other 
through the fluctuation-dissipation theorem 
\cite{vanKampen,Risken}. As it was clearly pointed out by 
Feynman in his {\it Lectures} \cite{Feynman}, the Second 
Law of Thermodynamics implies that a ratchet under the 
above conditions cannot generate current nor directional 
motion. Nevertheless, it is by now very well established that 
in order to obtain directional motion, the presence of an 
{\it external} unbiased time dependent force is necessary. 
This external force can be deterministic, a so-called 
``rocking'' ratchet \cite{Magnasco}, or stochastic 
in nature \cite{Doering}. In any case, any of those 
forces should be of zero average in time such that the 
current obtained is not a trivial consequence of the bias of 
those forces, but of their interplay with the ratchet potential. 
In the present article, since the external interaction of the 
protein during the folding process is supposed to yield energy 
through the consumption of high energy phosphate bonds, as
adenosine-triphosphate (ATP) or guanosine-triphosphate 
(GTP), we assume an external, unbiased, stochastic force.

We shall show that, understanding by protein folding the 
search and finding of a target point or small section in 
the energy landscape, the ratchet mechanism is extremely 
effective and robust. As mentioned above, an interesting
consequence of this process is that the landscape does 
not require a funnel structure \cite{Wolynes,Dill}.

\section{The model}

The model consists of an overdamped Brownian ``particle'' 
moving in a multidimensional configuration space, the 
coordinates representing the effective degrees of freedom 
of a protein \cite{note}. The environment enters as a 
dissipative thermal bath and the full folding process 
is driven by an unbiased external source.

\begin{figure}
\begin{center}
 \includegraphics[width=1.0\textwidth]{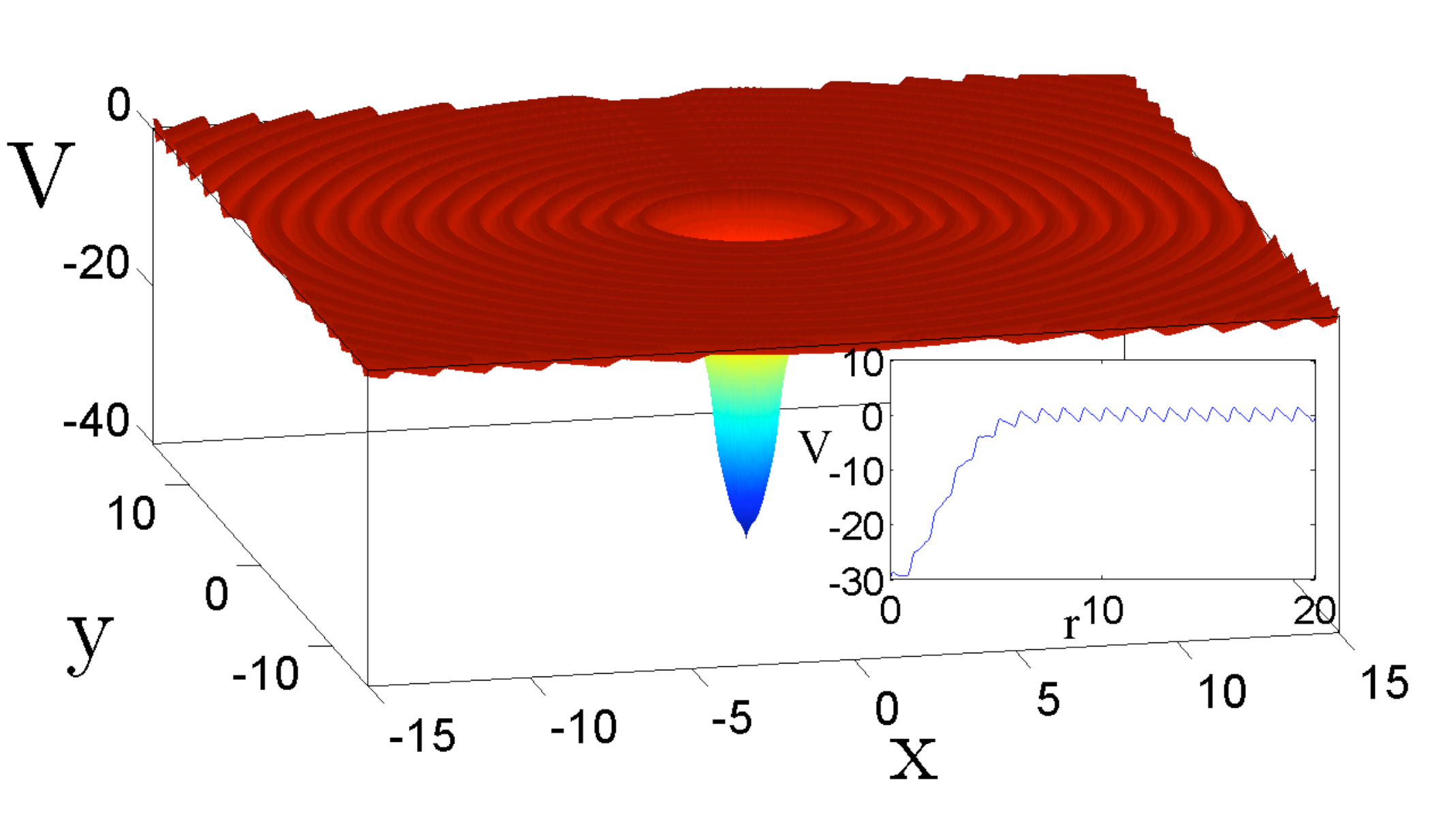}
\end{center}
\caption{(Color online)
Two-dimensional representation of the protein 
``free energy'' landscape, Eq.(\ref{V}), with $\lambda =1$, 
$V_0=1$, $V_1=50$, $a=0.4$, $b=0.1$ and $c=10$. This landscape is 
flat on the average and has a ratchet-like asymmetry {\it away} and 
{\it towards} the native state, represented by a deep and wide enough 
minimum of the potential.}
\label{landsimple}
\end{figure}

Our central proposal is that the regularity and intrinsic 
asymmetries found in polypeptides, such as $\alpha$ helices 
and $\beta$ sheets, may arise from or give rise to an asymmetry 
at the level of the energy landscape. This feature is introduced 
through an asymmetric ratchet-like, conservative potential. 
The potential may be flat on average but it is provided with 
a small, deep and wide enough well,
whose local minimum vicinity represents the native state of the 
protein. On top of the average potential there is a ratchet-like 
asymmetry that makes the potential {\it towards} the native state 
{\it different} than in the direction moving {\it away} from it. 
Figure \ref{landsimple} is a two-dimensional ``free energy'' 
landscape representation of the potential. Although the precise 
form of the landscape is essential for a given protein, here 
we shall use a very simple form for the configuration potential 
in order to show the virtues of the model. In $d$ dimensions 
the ratchet-like potential with a native state is given by, 
\begin{eqnarray}
	V(\vec{x}) &=& V_0\left[\sin\left(\frac{2\pi r}{\lambda}\right)
	+ a\sin\left(\frac{4\pi r}{\lambda}\right)
	+ b\sin\left(\frac{6\pi r}{\lambda}\right)\right]  \nonumber \\
        &&- V_1 e^{-r^2/c}, 
	\label{V}
\end{eqnarray}
where $\vec{x}=x_1,x_2,...,x_d$ are the configuration coordinates; 
$r=(x^2_1+x^2_2+...+x^2_d)^\frac{1}{2}$; 
and $a$, $b$, $c$, $V_0$, $V_1$ and $\lambda$ are positive real
numbers that determine the structure of the potential. In the 
figure captions we specify their particular values. The native 
state is arbitrarily set at $r = 0$ and it is a well given by 
the last term in $V(\vec x)$. The first three terms give 
rise to a periodic ratchet potential with a radial asymmetry.

The above landscape potential can be make extremely more 
complicated and still work in the same way as the stylized model. 
That is, nonperiodic ratchets with wells, barriers or 
other structures can be introduced in order to approximate the 
potential to the real conditions of proteins, without affecting 
the essence of the model. This will be shown below.

Proteins are typically embedded in a viscous environment at a 
fixed temperature. This is represented in the model by a thermal 
bath exerting both a dissipative and a white, Gaussian stochastic 
force. We believe this does not require further justification.  

With  the last two ingredients {\it only}, namely, with the 
ratchet potential and the thermal bath, the ``particle'' will 
obey Levinthal paradox and would not find its way to the native 
state, particularly in a multidimensional space. The directed 
transport mechanism requires the presence of a time-dependent 
{\it external} source of energy acting on the particle in order 
to produce work, in the form of a directional current \cite{Ibarra}. 
To represent a natural, non-designed process, this force must 
be unbiased. It turns out that the interplay of the asymmetry 
of the conservative ratchet potential with the external source 
of energy, yields the sough for directed motion. We shall 
demonstrate below how the introduction of this force makes 
the particle finds its way toward the native state and, on 
the average, in a very short time \cite{GCRR}. The efficacy 
of the process depends, of course, on the values of the 
different model parameters.

Regarding the origin of the external energy source, we 
generally argue that processes in  living organisms occur 
via non-equilibrium states with all sorts of gradients of 
different physical quantities, and that the maintenance 
of those gradients may be traced back to the consumption 
of energy. In a more specific way, during the protein synthesis, 
the action of the ribosome it is an important influence to the 
folding process, in the sense that this process is almost concluded 
in an average size protein by the time the ribosome releases its 
C-terminal end \cite{Alberts}. Furthermore, ribosomes 
are recognized complex molecular motors that consume 
energy in the form of GTP throughout the protein synthesis 
\cite{Sinha-Bhalla,Visscher,Garai}.
Although it seems it does not occur in all folding cases, 
molecular chaperones may represent another good example of 
an external ``force'' in the folding process since they 
consume ATP while assisting in the process 
\cite{Ellis,Hartl,Bhutani,Walter,Goloubinoff}.
In a broader sense, we may also argue that of all the many 
processes occurring in the crowded environment of the cell, 
some of them consume energy and can be used by the protein 
during the folding process as the required external force 
to reach the native state.

The mathematical model is, therefore, the overdamped Langevin 
equation for a Brownian particle embedded in a thermal bath 
moving in a multidimensional coordinate system, under the influence 
of a conservative potential $V(\vec{x})$ and in the presence of a 
time-dependent external force $F_i(t)$, 
\begin{equation}
    \gamma \frac{dx_i(t)}{dt} = f_i(t) 
    - \frac{\partial V(\vec{x})}{\partial x_i} + F_i(t).
    \label{eclan1}
\end{equation}
 with  
$i=1,2,\dots ,d$.  $\gamma$ is the 
friction coefficient of the bath; $f_i(t)$ is an stochastic Gaussian
thermal force, with zero mean and with a
white spectrum, namely\cite{vanKampen}:  
\begin{equation}
\langle f_i(t) f_j(t^{\prime}) \rangle = 2\gamma kT 
\delta (t-t^{\prime}) \delta_{ij},
\label{autocorr}
\end{equation}
with $i,j=1,2,\dots ,d$. In this expression $T$ is the temperature 
and $k$ Boltzmann constant. As mentioned, $\langle f_i(t) \rangle =0$ 
for all $i$. Thus, $\vec{f}(t)$ represents unbiased white 
noise. The fact that the thermal force correlation is proportional 
to $\gamma kT$ is the statement of detailed balance between the 
thermal bath and the particle \cite{vanKampen,Risken}.

As mentioned, the forces $F_i(t)$ represent the action of 
agents {\it external} to the protein itself and the thermal bath, 
and its driving is caused by ATP consumption. 
We really do not know the precise form of this force. However, 
as far as producing directed transport, its precise form is 
not essential, as long as its correlation time is different from 
that of the bath \cite{Magnasco,Ibarra}. Therefore, we have decided 
to employ a stochastic force generated by an Ornstein-Uhlenbeck 
process \cite{vanKampen,Risken} given by,
\begin{equation}
\frac{dF_i(t)}{dt}=-\frac{F_i(t)}{\tau _e}+\zeta_i(t),
\label{ecfe}
\end{equation}
for $i = 1, \dots, d$ and where the functions $\zeta_i(t)$ represent 
independent stochastic variables with zero mean 
$\left\langle \zeta_i(t)\right\rangle =0$, and Gaussian white noise, 
\begin{equation}
\left\langle\zeta_i(t)\zeta_j(t^{\prime })\right\rangle
=\frac{f_o^{2}}{\tau _e^{2}} \delta(t-t^{\prime}) \delta_{ij}. \label{zeta}
\end{equation}
The parameters $\tau_e$ and $f_o$ determine the process and 
represent the correlation time and the fluctuations magnitude 
of the external force, respectively. As we pointed out above, 
it has been proved that Brownian motors also work with different 
kind of zero bias forces, such as sinusoidal or dichotomic ones \cite{Reimann}.
 
\section{Results}

We solve Eqs.(\ref{eclan1})-(\ref{zeta}) numerically. We first 
fix the potential 
parameters, $V_0$, $V_1$, $a$, $b$, $c$ and $\lambda$, and then we 
vary $\gamma$, $T$, $f_o$ and $\tau _e$ such that the directed transport 
process is optimized. Due to numerical easiness, we study the evolution 
of a protein with six degrees of freedom. However, to show the robustness 
of the model we also present results for a protein with 100 degrees 
of freedom. This may represent a real protein with several tens of 
aminoacids.
 
\subsection{Folding in 6 dimensions}

\begin{figure}
\centering
\scalebox{0.5}{\includegraphics{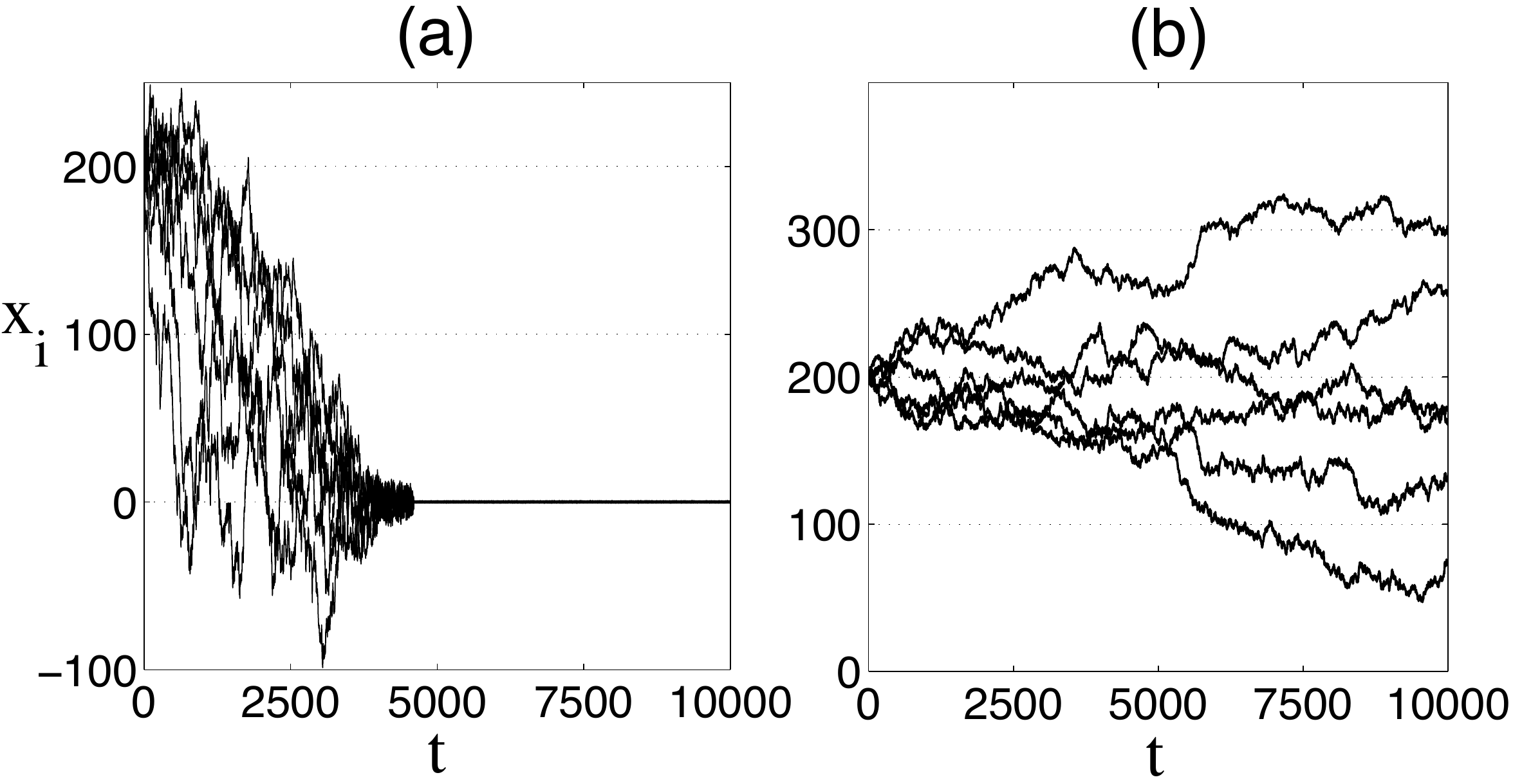}}
\caption{Typical folding realization of a protein with 6 degrees of 
freedom. Each line refers to the evolution of a single coordinate 
$x_i$, with $i = 1, \dots ,6$. (a) The stochastic external 
force is included. (b) No external force is considered and no folding 
occurs. In this case, 
$\gamma =1$, $kT=0.1$, $f_0=3.0$ and $\tau _e=0.5$. The initial 
condition was 200 units in each coordinate for both cases.}
\label{part_fe}
\end{figure}
Figures \ref{part_fe}a and  \ref{part_fe}b show typical examples of 
the particle evolution in a 6 dimensional space, by following the 
value of each of the coordinates $x_i$ for $i = 1, \dots, 6$. The 
behavior when the stochastic external force is present is shown in 
Fig. \ref{part_fe}a, while Fig. \ref{part_fe}b corresponds to 
the case when no external force is considered (and the temperature 
is increased in order to compensate in magnitude the lack of the 
external force). An evident directed transport towards the native 
state is observed in the former case, in contrast with the 
random-walker behavior when no external force is added. The directed 
transport occurs (Fig. \ref{part_fe}a) until the particle reaches the 
vicinity of the native state well. Then, a fluctuation takes place 
and the particle finally falls into the well. If the well is not wide 
enough the particle may stay for a long time around the minimum of the 
well. Once the particle is within the well, its position fluctuates 
depending on, the width of the well at its bottom, the magnitude of 
the two stochastic forces, and the dimensionality of the space. 
Typically, for spaces with higher dimensionality wider wells 
need to be considered in order to make the particle reach the 
native state. 
\begin{figure}
\centering
\scalebox{.5}{\includegraphics{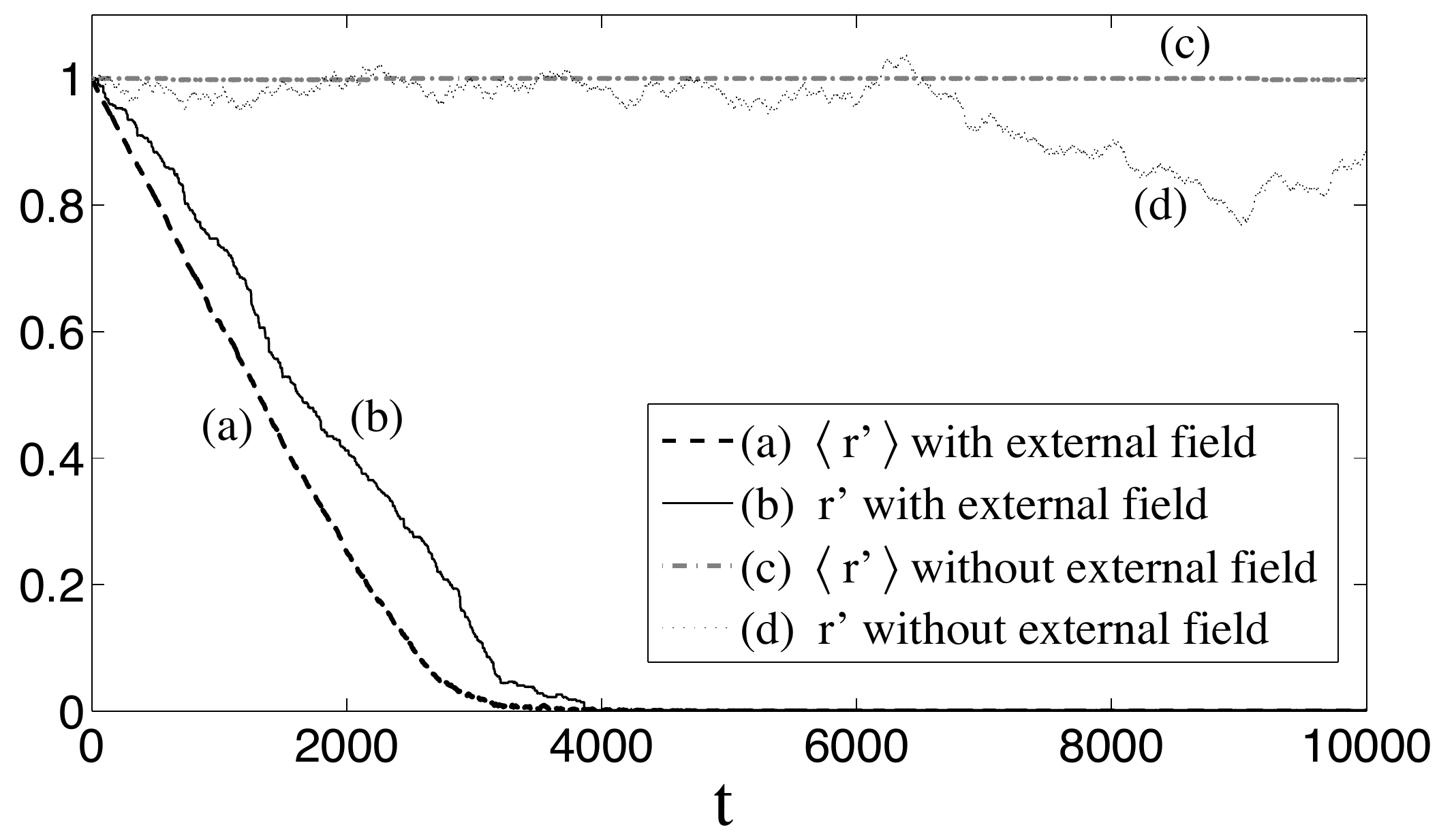}}
\caption{Time evolution of the dimensionless distance to the native state 
$r'=(x^2_1+x^2_2+...+x^2_d)^\frac{1}{2}/r_o$, with $r_o$ the 
initial condition, for a protein with 6 degrees of freedom. 
Single realizations and averages over 100 realizations are shown for both cases, 
with and without external forces. Same parameters as in Fig.\ref{part_fe}.}
\label{rsobs}
\end{figure}
In Fig. \ref{rsobs}, we show the evolution of the dimensionless radius 
position $r'=(x^2_1+x^2_2+...+x^2_d)^\frac{1}{2}/r_o$, with $r_o$ the 
initial condition. This figure corresponds to typical cases such as 
those analyzed above as well as the average for 100 realizations. 
For the relevant case when the external force is included, cases (a) 
and (b) in the figure, we have found 100\% efficacy of ``folding'' 
within the arbitrary time limit of 10000 units of time for the 
parameters considered. This time is very short even in CPU time scales. 
When the external force is absent, the particle is a true diffusing 
random walker and, insofar our numerical calculations, we have never 
seen even a single occurrence of 
finding the native state. This point serves to emphasize the role of 
the external force as the responsible one for delivering the energy 
to be further converted by the ratchet into useful work. The result 
of the work done is the folding of the protein.

\subsection{Robustness}
In order to demonstrate the robustness of the model two 
representative cases were studied. The first consisted on 
increasing the degrees of freedom up to 100 to verify that the 
dimensionality of the space has no fundamental influence on the behavior 
of the Brownian motor. The second case considers the inclusion 
of obstacles into the energy landscape, such as potential barriers 
and wells, in order to evaluate the capacity of the folding process 
to overcome such expected features in more realistic scenarios. 

\subsubsection{100 dimensions}

\begin{figure}
\centering\centering
\scalebox{.25}{\includegraphics{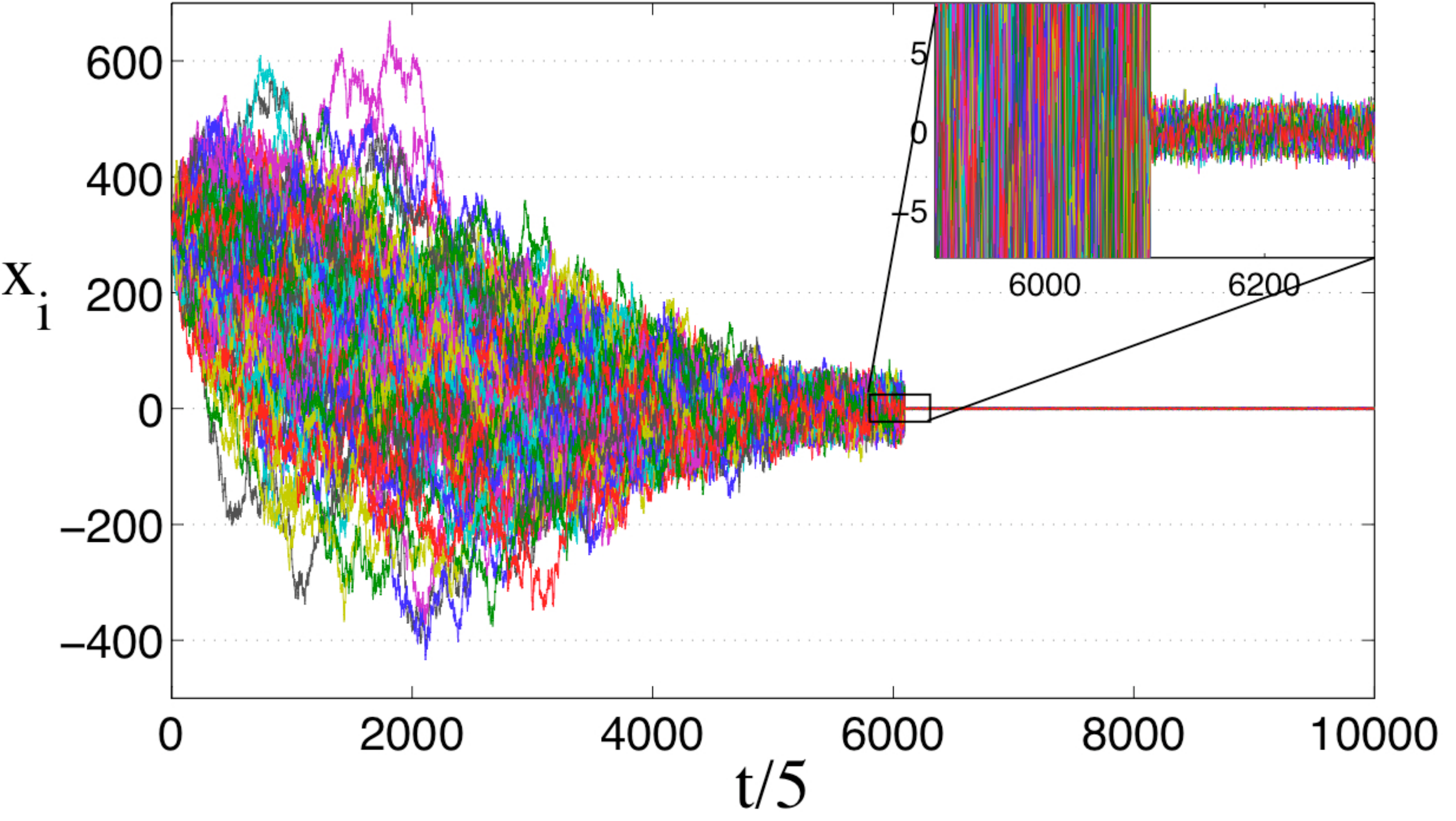}}
\caption{Typical folding realization of a protein with 100 degrees of 
freedom, with the external force present.  Each color refers to the 
evolution of a single coordinate $x_i$, with $i = 1, \dots ,100$. Here, 
we used the same potential parameters as in Fig.\ref{landsimple} except
$V_1 =10000$ and  $c=6000$, while the other parameters are: 
$\gamma =1$, $kT=0.1$, $f_0=3.0$ and $\tau _e=0.5$. 
The initial condition is 300 units for each coordinate.}
\label{r100d}
\end{figure}
\begin{figure}
\centering
\scalebox{.5}{\includegraphics{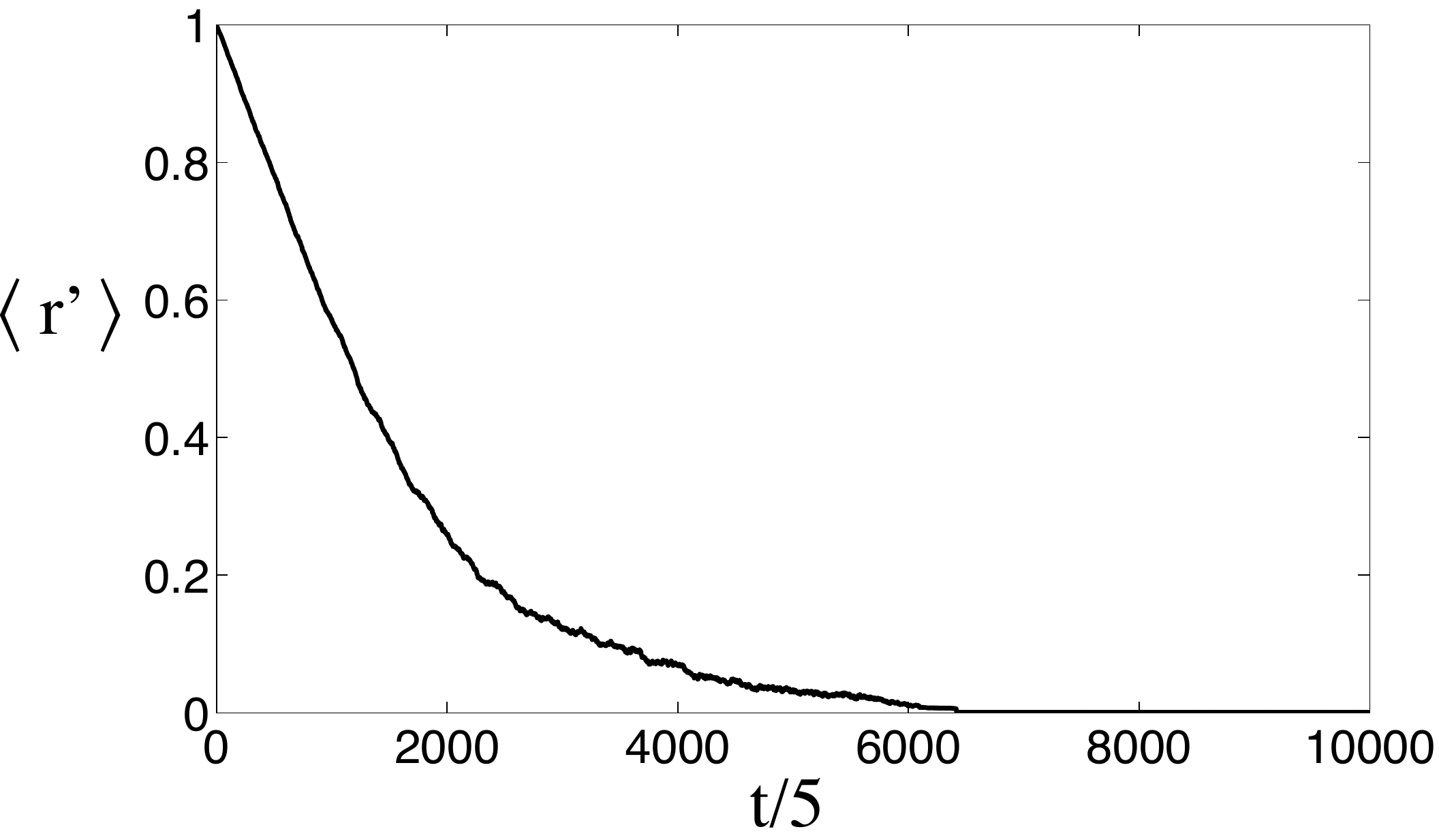}}
\caption{Time evolution of the dimensionless distance to the native state 
$r'$, see Fig.\ref{rsobs}, averaged over 20 realizations for a protein 
with 100 degrees of freedom. Same parameters as in Fig.\ref{r100d}.}
\label{r100dprom}
\end{figure}
Figure \ref{r100d} shows the evolution of the 100 state coordinates of 
a protein with 100 degrees of freedom. As before, the particle is under 
the influence of a potential such as that of Eq. (\ref{V}) and the external 
field is included. Figure \ref{r100dprom} shows the time dependence of the 
dimensionless radius $\langle r' \rangle$, averaged over 20 
realizations with the same conditions as the case shown in  
Fig. \ref{r100d}. In this case, the initial condition is 300 spatial 
units for each of the coordinates. We find that the overall behavior 
for 100 coordinates is essentially the same as for the 6-dimension case, 
showing an evident directed transport towards the native state with 
100\% of success reaching it. There are, as expected, quantitative 
differences as the dimensionality is increased. One is the increment 
in time to reach the native state, as seen in the figures. Other is 
the increase in size of the spatial fluctuations around the well. Both 
effects are a consequence of having much more space to explore in the 
directions orthogonal to the radial one, which is the unique direction where
the ratchet asymmetry is 
present. To be precise, within the parameters used, in order for 
the particle to get into the native state well, it was necessary to 
increase the width of the well from 10 to 250 space units as increasing 
from 6 to 100 dimensions. Further, in order to achieve small fluctuation 
amplitudes in the native state, see the zoom-in graph in 
Fig.\ref{r100d}, the width at the bottom of the well was kept narrow 
but its depth was increased. It should be also clear that the time to 
reach the native state depends on how far from it the particle is placed
initially. At the model level this is a matter of CPU time only.

\subsubsection{6 dimensions with obstacles}

\begin{figure}
\centering
\scalebox{.5}{\includegraphics{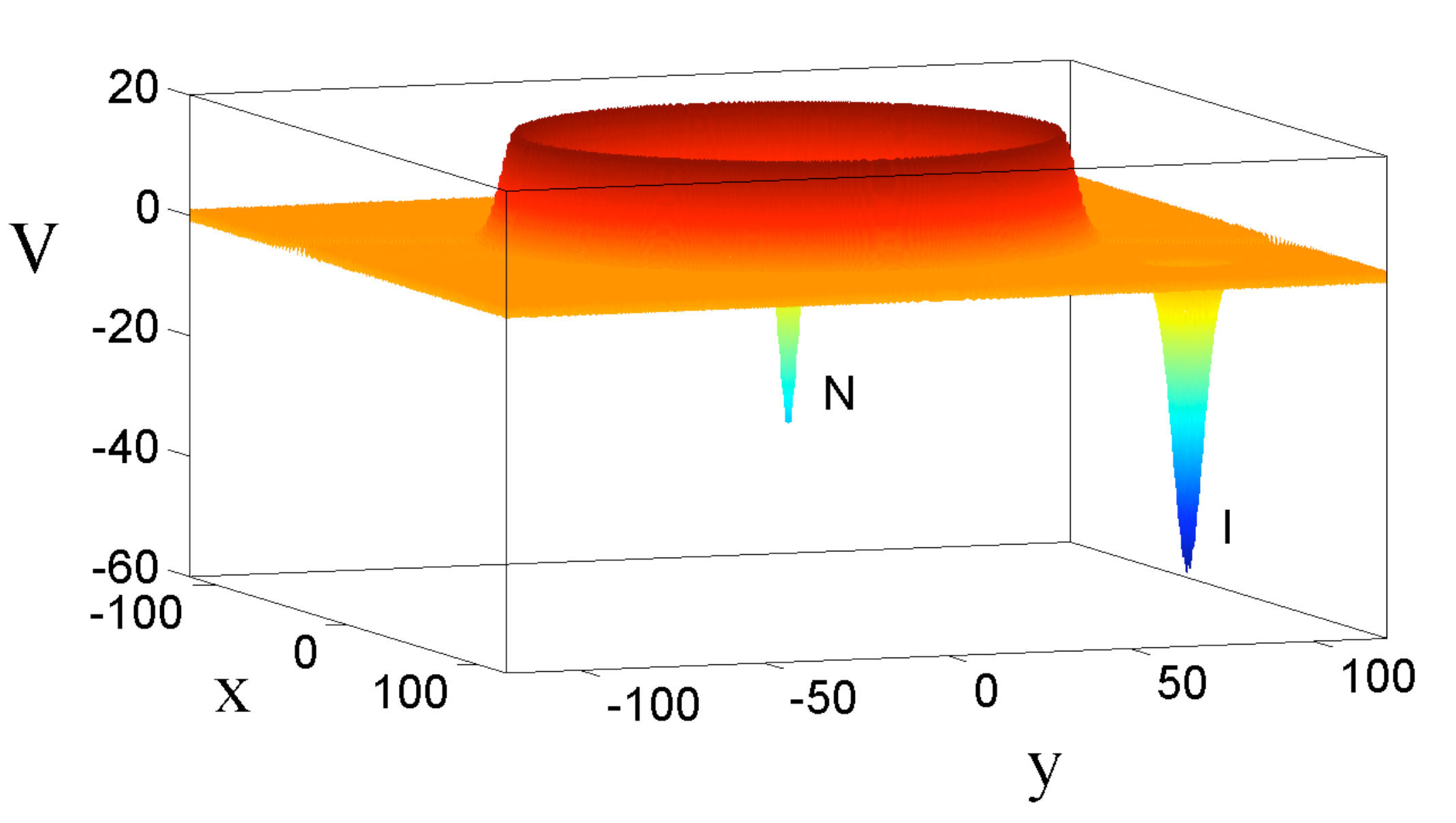}}
\caption{Bidimensional representation of the potential with obstacles, 
Eq. \ref{Vobs}. Here, $\lambda =1$, 
$V_0=1$, $V_1=50$, $a=0.4$, $b=0.1$, $c=10$, $V_b=18$, $V_w=60$, 
$c_b=30$, $c_w=30$, $r_b=250$, and $x^w_i=200$ for $i=1, \dots ,d$.
This conservative potential corresponds to the energy landscape 
of a protein with the native state (N) centered in the origin 
and two additional 
obstacles, a well (I) far away from the native 
state and a barrier between the two wells.}
\label{landobs}
\end{figure}
The bidimensional representation of the energy landscape of a protein 
with two additional obstacles is shown in Fig. \ref{landobs}. The 
obstacles consist on, one, an additional Gaussian well (I) {\it deeper} and 
{\it wider} than the native state well (N), (implying in principle 
a more stable state) and, two, a Gaussian barrier between the two 
wells and surrounding the native state. The equation corresponding to 
this potential with obstacles in $d$ dimensions is given by
\begin{eqnarray}
	V(\vec{x}) = V_0\left[\sin\left(\frac{2\pi r}{\lambda}\right)
	+ a\sin\left(\frac{4\pi r}{\lambda}\right)
	+ b\sin\left(\frac{6\pi r}{\lambda}\right)\right]  \nonumber \\
        - V_1 e^{-r^2/c} - V_b e^{-(r-r_b)^2/c_b} + V_w e^{-s_w^ 2/c_w}, 
	\label{Vobs}
\end{eqnarray}
with $s_w^2 = (x_1-x^w_1)^2+(x_2-x^w_2)^2+ \dots +(x_d-x^w_d)^2$. The 
new parameters $V_b$, $V_w$, $c_b$ and $c_w$ are the
positive real numbers that determine the shape
of the barrier and the well introduced, while $r_b$ and 
$\vec{x_w}= (x^w_1, x^w_2, \dots ,x^w_d)$ determine their positions, 
respectively.

\begin{figure}
\centering
\scalebox{.5}{\includegraphics{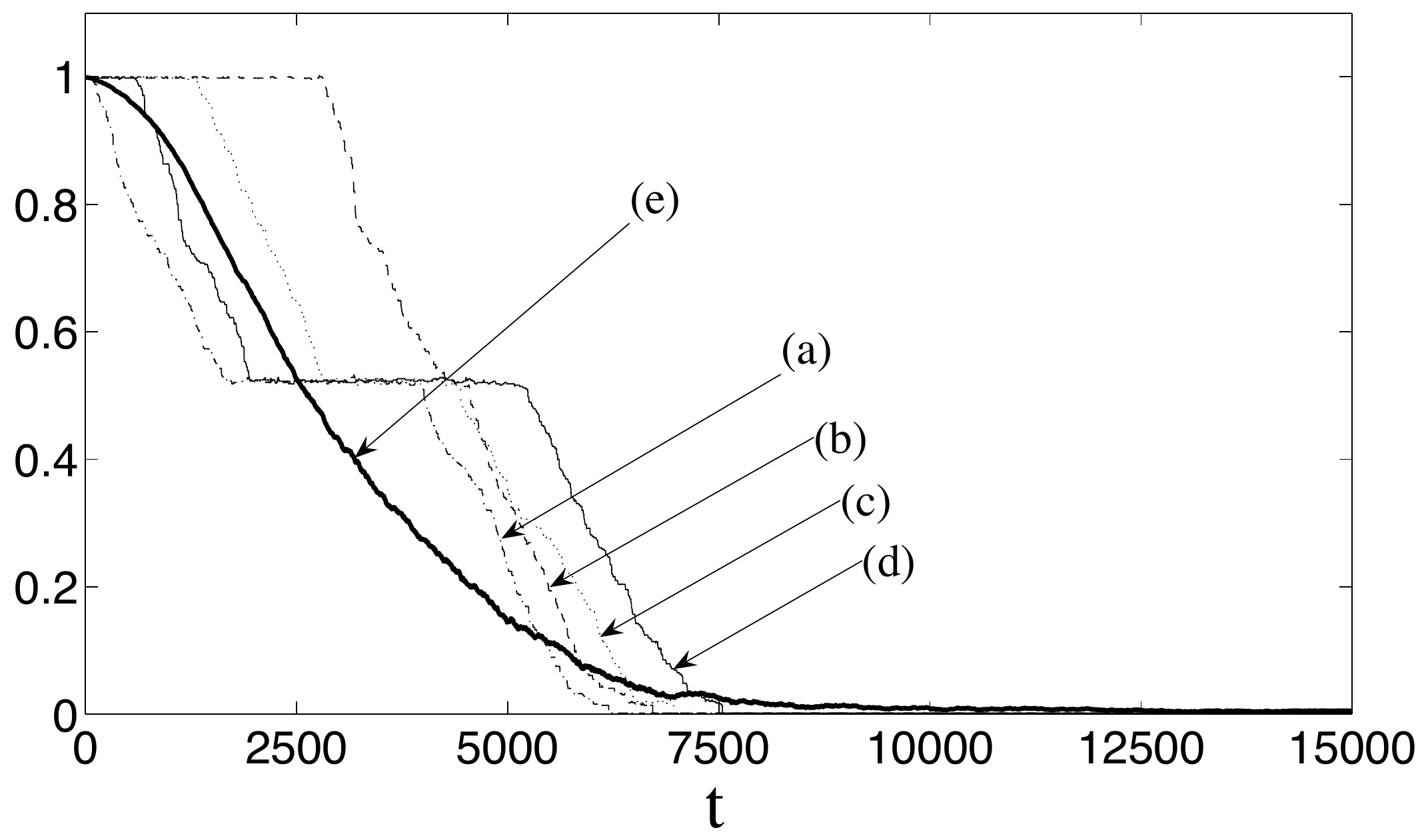}}
\caption{Time evolution of the dimensionless distance to the native state 
$r'$, see Fig.\ref{rsobs}, for a protein with 6 degrees of freedom 
in the potential with obstacles shown in Fig.\ref{landobs}. Cases (a)-(d) 
show typical single realizations and (e) is the average over 100 realizations.
The initial state is 200 units in each coordinate 
and coincides with the minimum of the additional well (I).}
\label{rcobs}
\end{figure}
The behavior of a particle in a 6 dimensional space, 
under the influence of the potential with obstacles, 
Eq. (\ref{Vobs}), and subjected to the stochastic external field, 
is shown in Fig. \ref{rcobs}. 
It is important to mention that the initial condition was set at 
the minimum of the added well (I). 
Fig. \ref{rcobs} shows that the particle is able to escape 
from the first, deeper but wider well, then overcome the potential 
barrier and finally find the native state. Frequently, the obstacles are 
not overcome immediately, as shown in the radius position evolution 
of typical realizations denoted by (a), (b), (c) and (d). We 
see regions of constant distances revealing that the particle takes 
some time leaving the first well and climbing the barrier. However, 
the particle always reaches the native state within the 
arbitrary time limit of 15000 time units, in an average time longer than 
the case with no obstacles , as the average radius 
of 100 realizations (e) illustrates.

Therefore, the protein ``folds'' even if severe obstacles, such
as barriers and potential wells, even deeper than that of the native 
state, are included in the energy landscape. This climbing capacity of 
the particle is another consequence of the delivery of work by the 
ratchet. The climbing capability does have a stall limit, thus, 
that depends on the local slope of the obstacle but not on its height 
or depth.

\section{Conclusions}

In this article we have revisited the ratchet or Brownian-motor 
mechanism extended to several dimensions in order 
to propose a model for protein folding.
The ratchet mechanism has become a potential candidate
for many biological processes, since, on the one hand
it resembles a thermal engine at a mesoscopic level, 
but on the other hand, it does not appear to need a 
specific design such as the man-made engines. 
That is, it needs only of two essential ingredients: an intrinsic
asymmetry and the input of external unbiased energy. The
result is delivery of work in the form of directional motion.

With this in mind several authors have used ratchets models 
in more obvious biological situations in which a transport process, 
of some kind, is present \cite{Reimann,Hanggi1,Bartussek,Astumian1}. 
Protein folding appears as a process with a directional 
motion towards the native state in the multidimensional 
space of the variables that define the energy landscape. 
These facts persuade us to propose a model for the 
mechanism of protein folding based on a Brownian motor.

Traditionally, this process has been thought to be driven 
by free energy differences leading towards a minimum, 
similarly to a chemical reaction \cite{Wolynes2}, and thus 
the idea of a funnel in the free energy landscape of the 
protein is very appealing. The contribution of external agents 
are welcome in the theory, however, they are not required.
Here, by following the idea that Life processes consume 
energy to yield their products, we have speculated that 
protein folding is driven by an ``engine'' in which the 
protein itself is part of it. 
Therefore, the presence of external sources of energy, 
provided by the ribosome, molecular choperones or other 
elements which consume energy during the process, becomes 
a necessity for the model presented here.

We have focused our work to the {\it in vivo} protein folding 
phenomenon since there are more evident possible sources for the 
required external energy. However, we think that the model presented 
can also be considered for the {\it in vitro} situation, where 
part of the bias and external energy necessary to obtain directional 
work to refold the proteins, may be provided by  the laboratory 
procedures, such as changes of temperature, pH or concentration 
of urea \cite{Lapanje,Auton}, among others.

On the other hand, at this moment the most difficult part to 
justify in our model is the asymmetric structure of the 
energy landscape. The best that we can say now is 
that the intrinsic asymmetric structure of proteins, 
which also presents certain periodicity, encourages us to 
think that ratchet-like structures can be hidden into the 
energy landscape.
Nevertheless, we do not have evidence neither pro nor con
that a ratchet structure is present since this would have to
be seen in the highly multidimensional energy landscape.

By means of numerical solutions of the model, it was 
found that the present model in 6 dimensions is able to 
produce directed transport in the mentioned space, 
through which the polypeptide is able to found the native 
state starting from an arbitrary initial state in the energy 
landscape and with a 100\% success. This happens when 
the external source of energy is present, otherwise the finding
of the native state is essentially never achieved. When the 
number of degrees of freedom is extended to 100, again, the 
process is performed with 100\% efficacy, but due to the 
larger number of dimensions, the time needed to reach the 
native state increases approximately 6-fold with
the parameters here used. On the other hand, 
the model demonstrates to be quite robust by showing that the
protein is able 
to overcome severe obstacles in the energy 
landscape in its way to the target state. 

To summarize and to emphasize it again, the interesting and 
powerful side of this model is that it achieves its goal by 
taking advantage of two simple 
requirements only: the intrinsic asymmetric properties of the 
system and the external energy source. The numerical 
solutions showed in this work prove that Levinthal's 
paradox can be solved with a model which does not require 
\emph{ad hoc} biased transition probabilities. In contrast, 
it requires the necessary driving of an external energy 
source, which in our opinion is physically appealing since its presence
is essential for Life to occur.


\begin{thebibliography}{36}
\bibitem{note} The multidimensional free energy landscape may 
already have the information that the system is immersed in water. 
That is, the interaction potential should have all effective 
interactions due to the hydrophilic or hydrophobic properties 
of the different aminoacids.
\end{thebibliography}
\end{document}